\documentclass[%
reprint,
amsmath,amssymb,
aps,
prb,
twocolumn,
]{revtex4-2}

\usepackage{color}

\usepackage{amssymb}
\usepackage{braket}
\usepackage{graphicx}
\usepackage{dcolumn}
\usepackage{bm}
\usepackage{hyperref}

\begin{document}

\title{Auxiliary Field Quantum Monte Carlo for Electron-Photon Correlation
}

\author{Braden M. Weight}
\affiliation{%
Theoretical Division, Los Alamos National Laboratory, Los Alamos, NM, 87545. 
}

\author{Yu Zhang}
\email{zhy@lanl.gov}
\affiliation{%
Theoretical Division, Los Alamos National Laboratory, Los Alamos, NM, 87545. 
}

\date{\today}

\begin{abstract}
Hybrid light-matter polaritonic states have shown great promise for altering already known and enabling novel chemical reactions and controlling photophysical phenomena. This field has recently become one of the most prominent and active areas of research that connects the communities of chemistry and quantum optics. The \textit{ab initio} modeling of such polaritonic phenomena has led to updating commonly used electronic structure methods, such as Hartree-Fock, density functional, and coupled cluster theories, to explicitly include Bosonic degrees of freedom. In this work, we explore the quantum electrodynamic auxiliary field quantum Monte Carlo (QED-AFQMC) method to accurately capture the polaritonic ground state of representative quantum chemical benchmark systems to explore electron-photon correlations. We analyze these correlations across multiple examples and benchmark the QED-AFQMC results against other \textit{ab initio} quantum electrodynamics methods, including QED-coupled cluster and QED–full configuration interaction, demonstrating the method’s accuracy and its potential for scalable simulations of strongly coupled light-matter systems.
\end{abstract}

\maketitle

\section{Introduction}
\label{sec: introduction}

Recent experimental studies of highly entangled light-matter states, known as polaritons, have been shown their ability to modify chemical reactions~\cite{Nagarajan2021JACS,GarciaVidal2021S,Hutchison2012ACIE,Schwartz2011PRL,Ebbesen2016ACR,Sau2021ACIE,Thomas2019S,Thomas2016ACIE,Thomas2020N,Lather2019ACIE,hirai_molecularPOL_ChemRev2023,Simpkins2023CR} and physical properties~\cite{Thomas2021,Vergauwe2019ACIE,Berghuis2020JPCC,Berghuis2022AP,xu_ultrafast_NatCommun2023,Deng2010RMP,rozenman_PolTransport_ACSPhot2018, Timmer:2023aa} of both low- and high-dimensional material systems,\cite{luttgens_population_2021,mohl_trion-polariton_2018,graf_near-infrared_2016,graf_electrical_2017,son_energy_2022,Allen_SWCNTpol_JPCC2022,mandal_microscopicQED_NanoLett2023,Lindoy2022JPCL} This has garnered a substantial interest in the theoretical community.\cite{ruggenthaler_QED_ChemRev2023, Mandal:2023cr, weight_QEDReview_PCCP2023, Foley:2023aa, Wang2021AP, haugland_GSQED_arXiv2023, Zhang2019JCP} Specifically, computational chemists have devoted the past few years toward the ``re-generation" of various many-body methods -- ubiquitously applied to pristine many-electron systems -- for use in the case of strongly correlated electron-photon systems.\cite{Foley:2023aa,weight_QEDReview_PCCP2023,Mandal:2023cr} All these approaches attempt to solve a non-relativistic quantum electrodynamical (QED) Hamiltonian for the coupled electron-nuclear-photon system for its eigenstates, usually within the cavity Born-Oppenheimer approximation~\cite{flick2017JCTC, weight_abQED_JPCL2023, Schnappinger:2023tx}, being the case for most conventional electronic structure methods. These modified electronic structure approaches include the QED Hartree-Fock (QED-HF)~\cite{haugland2020PRX, Schnappinger:2023tx, Cui:2024ul, riso2022NatComm, li_vQED_arXiv2023, Mazin:2024aa}, density functional theory (QED-DFT)~\cite{flick2018ACSPhotonics, pellegrini2015PRL, flick2018NanoPhotonics, Ruggenthaler2014PRA, Ruggenthaler2018NRC}, coupled-cluster theory (QED-CC) techniques~\cite{haugland2020PRX, Haugland2021JCP, Pavosevic2022JACS, deprince2021JCP, White:2020jcp}, quantum monte carlo (QMC)~\cite{Weight:2024pra, Weber:2025aa, Tang:2025aa},
and M{\o}ller-Plesset perturbation theory (QED-MP2)~\cite{Cui:2024ul}, to name a few. Moreover, the analogous methods for excited state simulations have also been recently developed, such as time-dependent QED-DFT (QED-TDDFT)~\cite{Flick2020JCP, yang2021JCP, yang_polGRADS_JCP2022, liebenthal2022JCP, liebenthal_mean-field_Arxiv2023, Vu_enhanced_JPCA2022}, equation of motion QED-CC (QED-EOMCCSD)~\cite{mordovina2020PRR, deprince2021JCP, liebenthal2022JCP}, and complete active space configuration interaction (QED-CASCI)~\cite{Vu:2024vv}.

In many of these cases, the drawbacks of the original method are exacerbated in its QED analog. For example, new exchange-correlation functionals need to be constructed to describe the electron-photon correlations in the DFT approach to account for such effects explicitly~\cite{flick2018ACSPhotonics, pellegrini2015PRL, Ruggenthaler2018NRC, Flick2018PRL, schafer2021PNAS, flick2022PRL}. The early attempts at developing such functionals resulted in a dramatic reduction in the quality of treatment of bare electron-electron correlations.\cite{weight_abQED_JPCL2023, Pavosevic2022JACS, Haugland2021JCP,liebenthal_mean-field_Arxiv2023, Vu_enhanced_JPCA2022} While efforts toward constructing improved functionals for the QED-DFT approach are on-going with marked success,\cite{schafer2021PNAS,flick2022PRL} the ``exact'' effects of the cavity presence on the electronic subsystem, are only trustworthy up to the choice of electron-photon and electron-electron exchange-correlation functionals.

While these methods have enabled valuable insights, they also inherit and amplify many of the limitations of their bare electronic counterparts. For example, QED-DFT relies critically on developing new exchange-correlation functionals capable of capturing light–matter interactions. Early functionals often introduced significant errors in electron–electron correlation. While more recent efforts have shown promise~\cite{schafer2021PNAS,flick2022PRL}, results still depend heavily on functional choice~\cite{weight_abQED_JPCL2023, Pavosevic2022JACS,Haugland2021JCP,liebenthal_mean-field_Arxiv2023, Vu_enhanced_JPCA2022}. Among wavefunction-based methods, QED-CCSD provides a systematic way to include electronic correlation on top of the QED-HF reference and has shown encouraging results for small molecules and moderate coupling strengths~\cite{haugland2020PRX, Pavosevic2022JACS}. However, due to the bosonic nature of the photon field and the need to truncate the number of photonic excitations, the method becomes inaccurate or intractable under strong coupling or for realistic molecular models~\cite{mordovina2020PRR}.

These theoretical challenges are especially pressing in light of rapid experimental progress. Modern plasmonic and dielectric nano-cavity designs have enabled access to the strong and ultra-strong coupling regimes under ambient or near-ambient conditions~\cite{akselrod_QEDNP_AdvMat2015, hoang_QDPlasmon_NanoLett2016}. These advances have motivated detailed studies of cavity-modified ground-state potential energy surfaces at high cavity frequencies (1–20 eV), revealing substantial changes in charge density~\cite{haugland2020PRX, deprince2021JCP, Weight:2024jacs}, intermolecular interactions~\cite{Haugland2021JCP}, and reaction barriers~\cite{Pavosevic2022JACS, Cui:2024ul, weight_abQED_JPCL2023}. Yet, an accurate and scalable method capable of treating strong electron–photon correlations without severe truncation is still lacking.

This work introduces a quantum electrodynamical auxiliary-field quantum Monte Carlo (QED-AFQMC) method for solving the Pauli-Fierz Hamiltonian with chemical accuracy and minimal approximations. Unlike conventional wavefunction-based approaches, AFQMC avoids explicit truncation of photonic excitations and naturally incorporates high-level electron–photon correlations via stochastic sampling. Building on the established success of AFQMC in fermionic quantum chemistry, we extend the method to treat both electronic and photonic degrees of freedom using a unified second-quantized framework. The resulting approach allows us to explore polaritonic ground-state properties and cavity-induced modifications to chemical energetics from first principles.

We demonstrate the capabilities of QED-AFQMC by first benchmarking against exact results for the HF molecule coupled to a quantized cavity mode, highlighting discrepancies between coupled-cluster-based approaches and the exact QED solution. We then apply the method to chemically relevant systems and isomerization reactions under strong light–matter coupling and analyze the impact of photonic excitations on potential energy surfaces and reaction barriers. Our results establish QED-AFQMC as a powerful new tool for accurately modeling polaritonic effects in molecules, with the potential to scale to larger and more complex systems.

\section{Theoretical Background}
\label{sec: theory}

\subsection{Hamiltonian}
\label{sec: hamiltonian}

\subsubsection{Molecular and Pauli-Fierz (PF) Hamiltonians}
The second-quantized many-electron Hamiltonian for real molecules within the Born-Oppenheimer approximation can be written as
\begin{equation}\label{EQ:H_EL}
    \hat{H}_\mathrm{e} = \sum^{M}_{pq} h_{pq} c^\dag_p c_q + \frac{1}{2} \sum^{M}_{pqrs} V_{pqrs} c^\dag_p c^\dag_q c_r c_s,
\end{equation}
where $M$ is the number of atomic orbitals (AOs),
\begin{equation}
    h_{pq} = \int d{\bf r} \phi_p^*({\bf r}) \hat{h} \phi_q({\bf r})
\end{equation}
and
\begin{equation}
    V_{pqrs} = \iint d{\bf r} d{\bf r}' \phi_p^*({\bf r}) \phi_q^*({\bf r}') \hat{V}_\mathrm{ee}(\boldsymbol{r},\boldsymbol{r}') \phi_r({\bf r}') \phi_s({\bf r})
\end{equation}
are the one-electron integral (OEI) and electron repulsion integral (ERI), respectively. The one-body operator is given by $\hat{h}(\mathbf{r}) = -\frac{1}{2} \nabla^2 + \hat{V}_{\text{ext}}(\mathbf{r})$, which includes both the kinetic energy and the external potential. The electron-electron repulsion operator is $\hat{V}_{\mathrm{ee}}(\boldsymbol{r}, \boldsymbol{r}') = \frac{1}{|\boldsymbol{r} - \boldsymbol{r}'|}$. As discussed later, we do not explicitly construct the ERI due to its prohibitive memory cost for large systems. Instead, we employ a block Cholesky decomposition~\cite{Henrik:2003rs, Nelson:1977si} to construct the Cholesky tensor, significantly reducing the memory footprint.

The molecular Hamiltonian can be extended to include electron-photon interactions via the Pauli-Fierz Hamiltonian,
\begin{align}
    \hat{H}_\mathrm{PF} &= \hat{H}_\mathrm{e} + \hat{H}_\mathrm{ph} + \hat{H}_\mathrm{e-ph} + \hat{H}_\mathrm{DSE},
\end{align}
where the photonic Hamiltonian $\hat{H}_\mathrm{ph}$, the bilinear coupling term $\hat{H}_\mathrm{e-ph}$, and the dipole self-energy $\hat{H}_\mathrm{DSE}$ are given by
\begin{align}
    \hat{H}_\mathrm{ph} &= \sum_\alpha^{N_p} \omega_\alpha \left(\hat{a}^\dag_\alpha \hat{a}_\alpha + \frac{1}{2}\right), \\
    \hat{H}_\mathrm{e-ph} &= \sum_\alpha \lambda_\alpha \sqrt{\frac{\omega_\alpha}{2}} \left({\bf \hat{e}}_\alpha \cdot \boldsymbol{\hat{\mu}}\right) \left(\hat{a}^\dag_\alpha + \hat{a}_\alpha\right), \\
    \hat{H}_\mathrm{DSE} &= \sum_\alpha \frac{\lambda_\alpha^2}{2} \left({\bf \hat{e}}_\alpha \cdot \boldsymbol{\hat{\mu}}\right)^2. \nonumber
\end{align}
Here, $\boldsymbol{\lambda}_\alpha = \sqrt{\frac{1}{\epsilon \mathcal{V}}} {\bf \hat{e}}_\alpha = \lambda_\alpha {\bf \hat{e}}_\alpha$ is the electron-photon coupling strength of the $\alpha^\mathrm{th}$ photon mode, with unit polarization direction ${\bf \hat{e}}_\alpha$, dielectric constant $\epsilon$ of the medium, and confining mode volume $\mathcal{V}$.
The molecular dipole operator includes both electronic and nuclear contributions:
\[
  \hat{\boldsymbol{\mu}} = -\sum_i^{N_\mathrm{e}} e {\bf \hat{r}}_i + \sum_I^{N_\mathrm{N}} Z_I {\bf R}_I,
\]
where $Z_I$ is the charge of the $I^\mathrm{th}$ nucleus

As shown in previous work, the DSE Hamiltonian depends solely on electronic operators, leading to modifications of the OEI and ERI. Consequently, the final Pauli–Fierz (PF) Hamiltonian can be rewritten as
\begin{align}\label{eq:finalpf}
    \hat{H}_\mathrm{PF} &= \sum_{pq}^{M} \tilde{h}_{pq}\hat{c}_p^\dag \hat{c}_q + \frac{1}{2}\sum_{pqrs}^{M} \tilde{V}_{pqrs} \hat{c}_p^\dag \hat{c}_q^\dag \hat{c}_r \hat{c}_s \\
    &\quad+ \sum_{pq}^{M}\sum_\alpha^{N_p} \sqrt{\frac{\omega_\alpha}{2}} g_{pq}^\alpha (\hat{a}^\dag_\alpha + \hat{a}_\alpha)\hat{c}_p^\dag \hat{c}_q \nonumber \\
    &\quad+ \sum_\alpha^{N_p} \omega_\alpha \left(\hat{a}^\dag_\alpha \hat{a}_\alpha + \frac{1}{2}\right), \nonumber
\end{align}
where $\tilde{h}_{pq} = h_{pq} - \frac{1}{2} \sum_\alpha^{N_p} Q_{pq}^\alpha$ and $\tilde{V}_{pqrs} = V_{pqrs} + \sum_\alpha^{N_p} g_{ps}^\alpha g_{qr}^\alpha$ are the DSE-modified one-electron and two-electron integrals, respectively. The coupling matrix is defined as $g^\alpha_{pq} = \bra{p} \lambda_\alpha (\boldsymbol{e}_\alpha \cdot \hat{\boldsymbol{\mu}}) \ket{q}$.
The term $Q^\alpha$ arises from the permutation of the electronic operators in the DSE Hamiltonian:
\begin{align}
  ({\bf \hat{e}}_\alpha \cdot \boldsymbol{\hat{\mu}})^2 & = \sum_{pqsr} g^\alpha_{ps} g^\alpha_{qr} \hat{c}^\dag_p \hat{c}_s \hat{c}^\dag_q \hat{c}_r \nonumber\\
  & = \sum_{pq\sigma\lambda} g^\alpha_{ps} g^\alpha_{qr} (\delta_{\nu\sigma} \hat{c}^\dag_p \hat{c}_r + \hat{c}^\dag_p \hat{c}^\dag_q \hat{c}_r \hat{c}_s). \nonumber
\end{align}
Note that the exact form of $Q^\alpha$ depends on the specific treatment of the DSE operator~\cite{Foley:2023aa, li_vQED_arXiv2023}. In this work, we assume a complete basis set, and therefore $Q^\alpha_{pq} = - \sum_{s} g^\alpha_{ps} g^\alpha_{sp}$.
Nevertheless, the final Pauli-Fierz (PF) Hamiltonian consists of the bare electronic and photonic Hamiltonians, along with a bilinear light-matter coupling term.
Although the DSE term originates from the transverse component of the electric field and may vanish in certain cavities (leading to the disappearance of the DSE-mediated contributions to the OEI and ERI in Eq.~\ref{eq:finalpf}), the formalism presented in Eq.~\ref{eq:finalpf} is general and remains valid regardless of whether the DSE is included.

\subsubsection{Monte Carlo (MC) Hamiltonian}
As shown later, the Auxiliary-Field Quantum Monte Carlo (AFQMC) formalism requires rewriting the original Hamiltonian into the format of a so-called Monte Carlo Hamiltonian,
\begin{equation}
    \hat{H}_{\text{MC}} = \hat{T} + \frac{1}{2} \sum^{N_\gamma}_\gamma \hat{L}^2_\gamma + C,
\end{equation}
which consists of a one-body operator $\hat{T} = \sum_{pq} T_{pq} \hat{c}^\dag_p \hat{c}_q$, squares of one-body terms $\hat{L}_\gamma = \sum_{pq} L_{pq} \hat{c}^\dag_p \hat{c}_q$, and a constant shift $C$. In the following, we discuss how to transform each component of the PF Hamiltonian into the MC formalism.

\textbf{Photonic Hamiltonian.} In this work, we only consider the non-interacting photonic subsystem. Therefore, the photonic Hamiltonian includes only the one-body term, as shown in Eq.~\ref{eq:finalpf}, and no additional pre-processing is needed. However, it is worth noting that extending the Hamiltonian to include photonic interactions is straightforward, as demonstrated in previous extensions of AFQMC to many-boson systems~\cite{Purwanto:2004uj}.

\textbf{Electronic Hamiltonian.} As shown in Eq.~\ref{eq:finalpf}, the electronic part of the PF Hamiltonian is formally the same as the bare electronic Hamiltonian, although the OEI and ERI are modified by the DSE. This work employs the modified Cholesky decomposition method~\cite{Shi:2021aa, Nelson:1977si, Henrik:2003rs} to construct the MC Hamiltonian while avoiding the explicit construction of the full (memory-intensive) ERI.
Using the modified Cholesky decomposition, the ERI is rewritten as
\begin{equation}
    \tilde{V}_{pqrs} = \sum_\gamma L^e_{pq,\gamma} L^{e,*}_{rs, \gamma}.
\end{equation}

Hence, the DSE-modified electronic Hamiltonian $\hat{\mathcal{H}}_e$ can be rewritten as
\begin{align}\label{eq:eri_reorg}
  \hat{\mathcal{H}}_{\text{e}} = & \sum^M_{pq} \tilde{h}_{pq} \hat{c}^\dag_p \hat{c}_q + \frac{1}{2} \sum^M_{pqrs, \gamma} L^e_{pq, \gamma} L^{e,*}_{rs, \gamma} \hat{c}^\dag_p \hat{c}^\dag_q \hat{c}_r \hat{c}_s \\
  = & \sum^M_{pq} \left(\tilde{h}_{pq} - \frac{1}{2} \sum_r L^e_{pr, \gamma} L^{e,*}_{qr, \gamma} \right) \hat{c}^\dag_p \hat{c}_q \nonumber \\
  & + \frac{1}{2} \sum_\gamma
  \left( \sum_{pq} L^e_{pq, \gamma} \hat{c}^\dag_p \hat{c}_q \right)
  \left( \sum_{rs} L^{e,*}_{rs, \gamma} \hat{c}^\dag_r \hat{c}_s \right).
  \nonumber
\end{align}

\textbf{Bilinear Coupling Term.}
For simplification, we rewrite the bilinear coupling in a general form $\hat{H}_{eb} = \sum_\alpha \hat{F}_\alpha \hat{B}_\alpha$. One way to rewrite this bilinear coupling term in the format of the MC Hamiltonian is via the identity
\begin{equation}\label{eq:decouple1}
  \hat{F}_\alpha \hat{B}_\alpha = \frac{(\hat{F}_\alpha + \hat{B}_\alpha)^2 - (\hat{F}_\alpha - \hat{B}_\alpha)^2}{4}.
\end{equation}

Hence, the Hubbard-Stratonovich (HS) transformation of the bilinear coupling term results in $2N_\alpha$ auxiliary fields, specifically $\left\{ \frac{\hat{F}_\alpha \pm \hat{B}_\alpha}{\sqrt{2}} \right\}$.

As an alternative treatment, the bilinear term can be decoupled using the identity:
\begin{equation}
    \hat{F}_\alpha \hat{B}_\alpha = \frac{A}{2} \left[ (\hat{F}_\alpha + \hat{B}_\alpha)^2 - \hat{F}_\alpha^2 - \hat{B}_\alpha^2 \right],
\end{equation}
which leads to $3N_\alpha$ auxiliary fields.
Depending on the choice of the $\hat{F}_\alpha$ and $\hat{B}_\alpha$ operators, each decomposition scheme leads to a different effective MC Hamiltonian and can exhibit different levels of fluctuation in importance sampling. For example, possible choices include:
$\{\hat{F}_\alpha = \lambda_\alpha \boldsymbol{e}_\alpha \cdot \hat{\boldsymbol{\mu}}$, $\hat{B}_\alpha = \sqrt{\frac{\omega_\alpha}{2}}(b^\dag_\alpha + b_\alpha)\}$,
and
$\{\hat{F}_\alpha = \sqrt{\frac{\omega_\alpha}{2}} \lambda_\alpha \boldsymbol{e}_\alpha \cdot \hat{\boldsymbol{\mu}}$, $\hat{B}_\alpha = (b^\dag_\alpha + b_\alpha)\}$.
Since the second decomposition introduces $3N_\alpha$ auxiliary fields, it generally results in larger fluctuations during sampling. When the number of photon modes $N_\alpha$ is small relative to the number of auxiliary fields arising from the Cholesky decomposition of the electron-electron interaction tensor, the difference in statistical efficiency between the two schemes is minimal. However, as $N_\alpha$ increases and becomes comparable to or exceeds the number of electronic auxiliary fields, the overhead introduced by the second scheme becomes significant. In such cases, the first decomposition scheme (with only $2N_\alpha$ auxiliary fields) is preferred due to its reduced fluctuations and improved numerical stability.

Nevertheless, after incorporating the electron-photon interaction Hamiltonian and reorganizing the one-body and two-body interactions, the final MC form of the original PF Hamiltonian reads
\begin{align}\label{eq:mcHam}
  \hat{H}_{\text{MC}}
  = & \hat{T} + \frac{1}{2} \sum^{N_\gamma}_{\gamma} \hat{L}^{e,2}_\gamma
  + \frac{1}{2} \sum^{N_{\gamma'}}_{\gamma'} \hat{L}^{ep,2}_{\gamma'}
  + \hat{H}_{\text{ph}}
  + C \nonumber \\
  & \equiv \hat{T} + \frac{1}{2} \sum^{\mathcal{N}_\gamma}_{\gamma} \hat{\mathcal{L}}^2_\gamma
  + \hat{H}_{\text{ph}}
  + C.
\end{align}
Here, $\hat{\mathcal{L}} \equiv \{\hat{L}^e, \hat{L}^{ep}\}$ includes the intrinsic bare electronic operators $\hat{L}^e$ and the $\hat{L}^{ep}_{\gamma'}$ operators resulting from the decomposition of the bilinear coupling term. The number of such operators, $N_{\gamma'}$, is either $2N_\alpha$ or $3N_\alpha$, depending on the decomposition scheme. For example, $\hat{L}^{ep}_{\gamma'} \in \left\{ \frac{\hat{F}_\alpha + \hat{B}_\alpha}{2}, \frac{i(\hat{F}_\alpha - \hat{B}_\alpha)}{2} \right\}$ for the decomposition scheme in Eq.~\ref{eq:decouple1}.
The effective kinetic operator is given by $\hat{T} = \sum_{pq} T_{pq} \hat{c}^\dag_p \hat{c}_q$, where
\begin{equation}
    T_{pq} = \tilde{h}_{pq}
    - \frac{1}{2} \sum_{\lambda r} L^{e}_{\gamma,pr} L^{e,*}_{\gamma,qr},
\end{equation}
which results from the reorganization of the two-body operators, as shown in Eq.~\ref{eq:eri_reorg}.

\subsection{AFQMC Scheme for Molecular Quantum Electrodynamics}

AFQMC computes the ground state via imaginary-time evolution:
\begin{equation}
    \ket{\Psi_0} \propto \lim_{\tau \rightarrow \infty} e^{-\tau \hat{H}} \ket{\Psi_T}.
\end{equation}
The ground state $\ket{\Psi_0}$ of a many-body Hamiltonian \(\hat{H}\) can be projected from any trial wavefunction \(\ket{\Psi_T}\) that satisfies the non-orthogonality condition \(\braket{\Psi_T | \Psi_0} \neq 0\). In this work, we construct the trial wavefunction as a direct product of electronic and photonic components, i.e., \(\ket{\Psi_T} = \ket{\Psi_T^e} \otimes \ket{\Psi_T^{\text{ph}}}\). Specifically, we use a Hartree-Fock determinant for the electronic part \(\ket{\Psi_T^e}\) and Fock states \(\{\ket{n}\}\) for the photonic component $\ket{\Psi_T^{\text{ph}}}$.
In practice, the imaginary-time evolution is discretized into a sequence of small time steps:
\begin{equation}\label{eq:itestep}
    \ket{\Psi^{(n+1)}} = e^{-\Delta\tau \hat{H}} \ket{\Psi^{(n)}},
\end{equation}
and the ground state is projected out in the limit \(n \rightarrow \infty\).
The choice of trial wavefunction plays a critical role in determining both the accuracy and efficiency of AFQMC simulations. It guides importance sampling, reduces variance, and directly affects systematic errors in constrained or phaseless variants of AFQMC. To improve accuracy, more sophisticated trial states—such as multi-reference wavefunctions for the electronic part and variational Gaussian ans\"atze (e.g., displacement~\cite{li_vQED_arXiv2023} or squeezing states~\cite{Mazin:2024aa}) for the photonic part—are expected to provide better overlap with the true ground state and accelerate convergence. The exploration of such advanced trial wavefunctions is the focus of future work.

When $\Delta\tau$ is sufficiently small, the one-body and two-body terms in the evolution operator can be further factorized by the Suzuki–Trotter decomposition~\cite{Suzuki:1976aa}. There are different Suzuki–Trotter decomposition schemes. In this work, we employ the widely used symmetric decomposition:
\begin{align}
    e^{-\Delta\tau \hat{H}_{\text{MC}}} &\approx e^{-\frac{\Delta\tau}{2} \hat{T}}
    e^{-\frac{\Delta\tau}{2} \hat{H}_{\text{ph}}}
    e^{-\Delta\tau \sum_\gamma \frac{\hat{\mathcal{L}}^2_\gamma}{2}}
    \nonumber \\
    & \quad \times e^{-\frac{\Delta\tau}{2} \hat{H}_{\text{ph}}}
    e^{-\frac{\Delta\tau}{2} \hat{T}} e^{-\Delta\tau C} + \mathcal{O}(\Delta \tau^3).
\end{align}

\subsubsection{Hubbard-Stratonovich Transformation: Auxiliary Fields}

The two-body propagators, $e^{-\Delta\tau \sum_\gamma \frac{\hat{\mathcal{L}}^2_\gamma}{2}}$,
can be decomposed into one-body propagators via the Hubbard–Stratonovich (HS) transformation:
\begin{equation}
    e^{-\Delta\tau \sum_\gamma \hat{\mathcal{L}}^2_\gamma / 2}
    = \prod_\gamma \int dx_\gamma \frac{1}{\sqrt{2\pi}}
    e^{-x^2_\gamma / 2} e^{x_\gamma \sqrt{-\Delta\tau} \hat{\mathcal{L}}_\gamma},
\end{equation}
where $\{x_\gamma\}$ are the auxiliary fields.
In other words, the original imaginary-time propagation involving complex two-body operators is transformed into a high-dimensional integral over one-body propagators, parameterized by the auxiliary fields, which can be efficiently evaluated using Monte Carlo techniques~\cite{Shi:2021aa}.
Regarding the bilinear term, the HS transformation of $e^{-\Delta\tau \sum_{\gamma'} \frac{\tilde{L}^2_{\gamma'}}{2}}$ leads to the decoupled propagation of fermionic and photonic components, as they commute:
\begin{equation}
  e^{x \frac{\hat{F}_\alpha \pm \hat{B}_\alpha}{\sqrt{2}}} =
  e^{x \hat{F}_\alpha / \sqrt{2}} e^{\pm x \hat{B}_\alpha / \sqrt{2}}.
\end{equation}
Note that although we decouple the electronic and photonic degrees of freedom, they remain indirectly coupled via the shared auxiliary fields.
In practice, we implement importance sampling by modifying the underlying probability distribution of the auxiliary fields, analogous to the biased sampling approach widely used in conventional AFQMC for fermion-only systems~\cite{Shi:2021aa}. Specifically, we shift the center of the Gaussian probability distribution to account for the force bias derived from the trial wavefunction, thereby reducing fluctuations and improving sampling efficiency.

\section{Numerical examples}
\label{sec: numerical}

\subsection{Computational details}

We use stochastic reconfiguration (SR) for population control in the AFQMC simulations. The mean and standard deviation of all observables are computed via autocorrelation analysis~\cite{Purwanto:2004uj}. A time step of $\Delta\tau = 0.005$ is used for imaginary time propagation, with a total propagation time of $T = 20$. We evaluate the energy estimator every 10 time steps in order to reduce computational cost. This is necessary because energy estimation involves a four-index contraction over two-electron integrals and thus scales as $\mathcal{O}(N^4)$ with system size, making it one of the primary computational bottlenecks in the simulation. Unless explicitly stated otherwise, all calculations use the 6-31G basis set for the electronic wavefunction and a truncated photonic Fock space including up to five photon number states. 

\subsection{Results}
In this study, we consider the hydrogen fluoride (HF) molecule to be a prototypical small diatomic system. Its modest size allows us to perform exact full configuration interaction (FCI) calculations, which serve as a high-accuracy benchmark for testing our quantum Monte Carlo approaches under varying light-matter interaction strengths. Additionally, the polar nature of HF makes it particularly sensitive to dipole-related interactions, such as the dipole self-energy (DSE), making it an ideal test case for cavity quantum electrodynamics (QED) effects.

\begin{figure}
    \centering
    \includegraphics[width=0.99\linewidth]{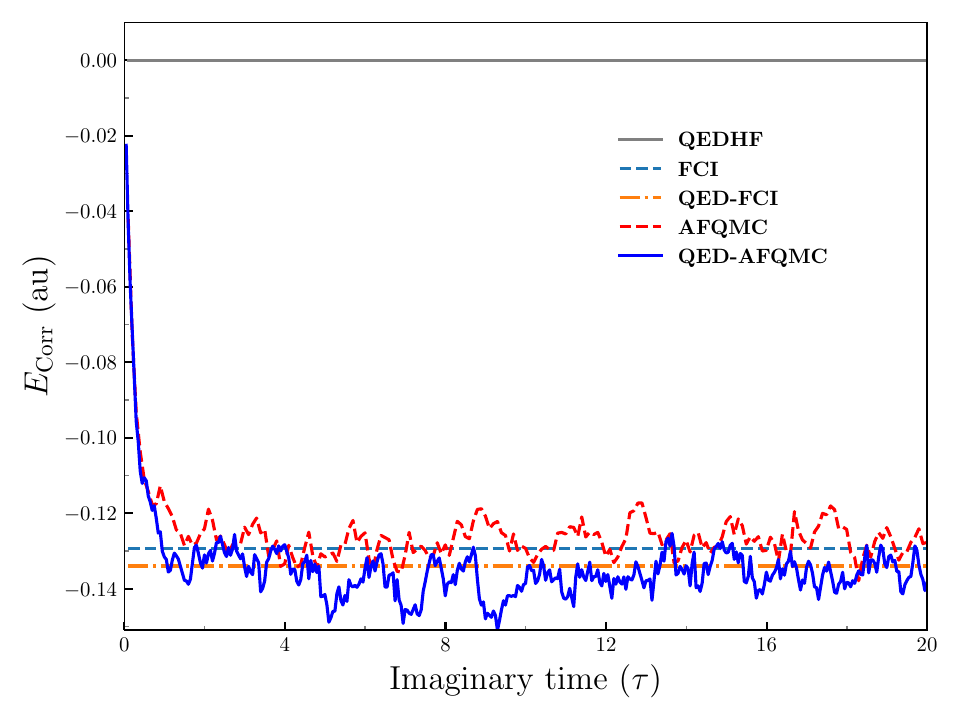}
    \caption{Quantum Imaginary time evolution within the AFQMC and QED-AFQMC schemes for the HF molecule with coupling strength $\lambda=0.1$. The only difference in the two calculations is the inclusion of photonic and electron-photon bilinear terms. }
    \label{fig1}
\end{figure}

In Figure~\ref{fig1}, we show the evolution of AFQMC energy as a function of imaginary time for an electron-photon coupling strength $\lambda=0.1$, comparing two variants of the method. The ``bare" AFQMC curve corresponds to simulations where only the electronic part of the PF Hamiltonian (with the DSE term) is propagated, while the QED-AFQMC results include the full light-matter Hamiltonian—namely, the photon mode and the bilinear electron-photon coupling term. 
The energy exhibits stable and smooth convergence in both cases, and the QED-AFQMC energy asymptotically approaches the FCI benchmark.
This demonstrates the numerical stability and accuracy of the QED-AFQMC method while confirming that incorporating photonic degrees of freedom does not compromise the convergence behavior of the underlying AFQMC algorithm.

The convergence profiles in Figure~\ref{fig1} validate the fidelity of our QED-AFQMC implementation. The fact that QED-AFQMC reaches the same ground-state energy as FCI, within statistical uncertainty, illustrates its capability to capture the essential quantum correlations introduced by both the electronic and photonic components. Furthermore, the agreement between bare AFQMC and FCI in the absence of explicit photon coupling (but with DSE included) supports our treatment of the fermionic part of the Hamiltonian. These results lay a solid foundation for using AFQMC-based approaches to explore more complex molecules or larger photonic Hilbert spaces, where FCI methods becomes computationally infeasible.

\begin{figure}
    \centering
    \includegraphics[width=0.99\linewidth]{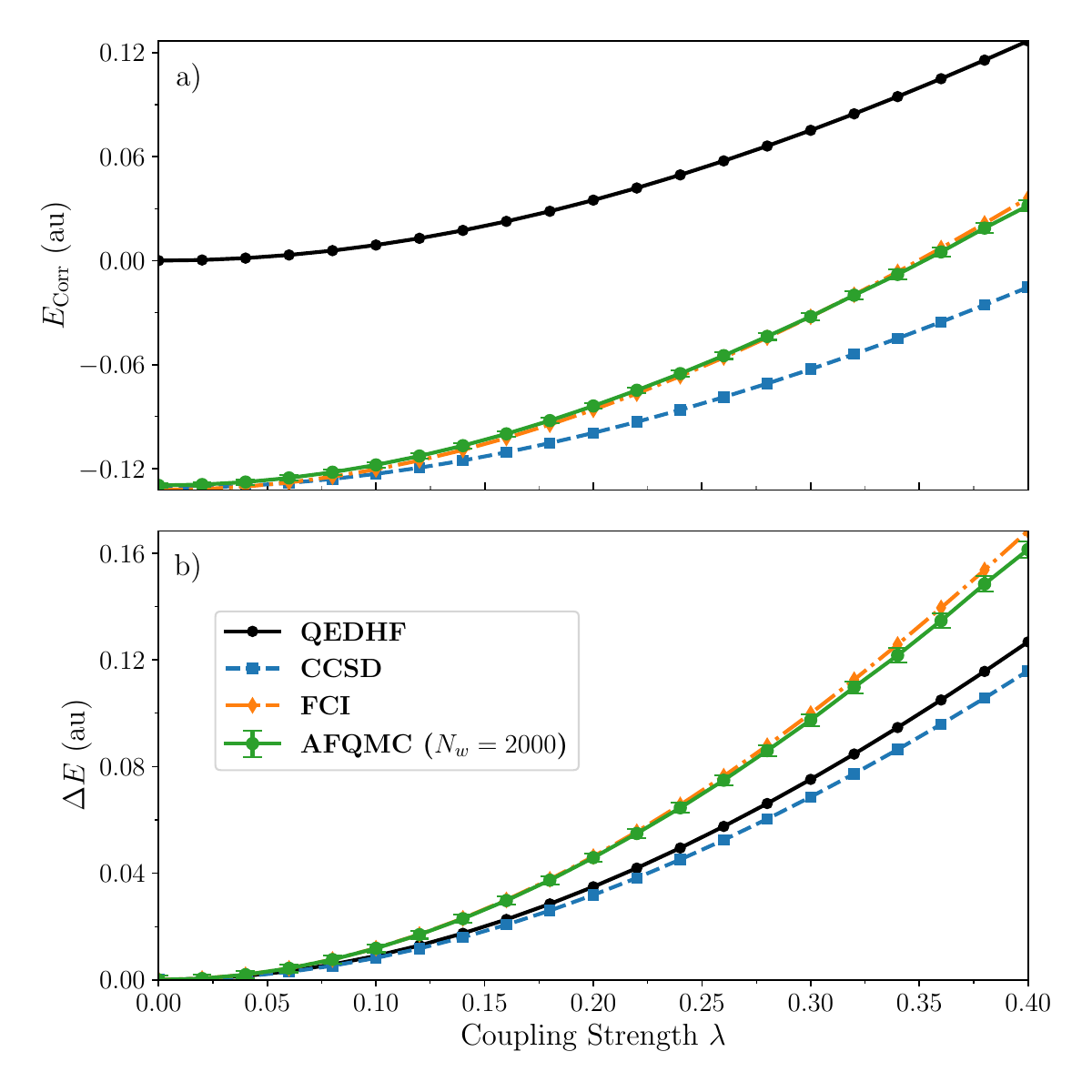}
    \caption{a) Correlation energies and b) DSE-mediated changes in correlation energies of the HF molecule as a function of electron-photon coupling strength $\lambda$ are compared using CCSD, FCI, and AFQMC. Here, we focus solely on the electronic subsystem of the PF Hamiltonian (including the dipole self-energy term). AFQMC exhibits excellent agreement with FCI across all $\lambda$, while CCSD systematically underestimates the correlation energy, particularly at stronger coupling strengths.
    }
    \label{fig2}
\end{figure}

In Figure~\ref{fig2}, we present a) correlation energies and b) DSE-mediated changes in correlation energies of the HF molecule computed at various electron-photon coupling strengths $\lambda$, focusing solely on the electronic subsystem of the PF Hamiltonian (including the dipole self-energy term). Since the photon degrees of freedom are excluded here, the Hamiltonian remains purely fermionic, allowing us to compare methods from conventional quantum chemistry directly. We compute the energies using HF, CCSD, FCI, and AFQMC. Across the entire $\lambda$ range, AFQMC and FCI are in excellent agreement, while CCSD systematically underestimates the correlation energy, and HF yields consistently higher energies due to its lack of correlation.
The comparison in Figure~\ref{fig2} highlights the importance of using accurate many-body techniques in the presence of the DSE, even when the photon field is not explicitly treated. As the coupling strength increases, the limitations of CCSD become more pronounced, likely due to its truncation at the doubles excitations and inability to fully capture DSE-induced correlations. AFQMC, by contrast, maintains its accuracy across the coupling regime and continues to track the exact FCI results, demonstrating its robustness and scalability for future studies in cavity-modified chemistry and quantum optics. These findings validate AFQMC as a high-fidelity solver for both electronic and light-matter hybrid Hamiltonians.

\begin{figure}
    \centering
    \includegraphics[width=0.99\linewidth]{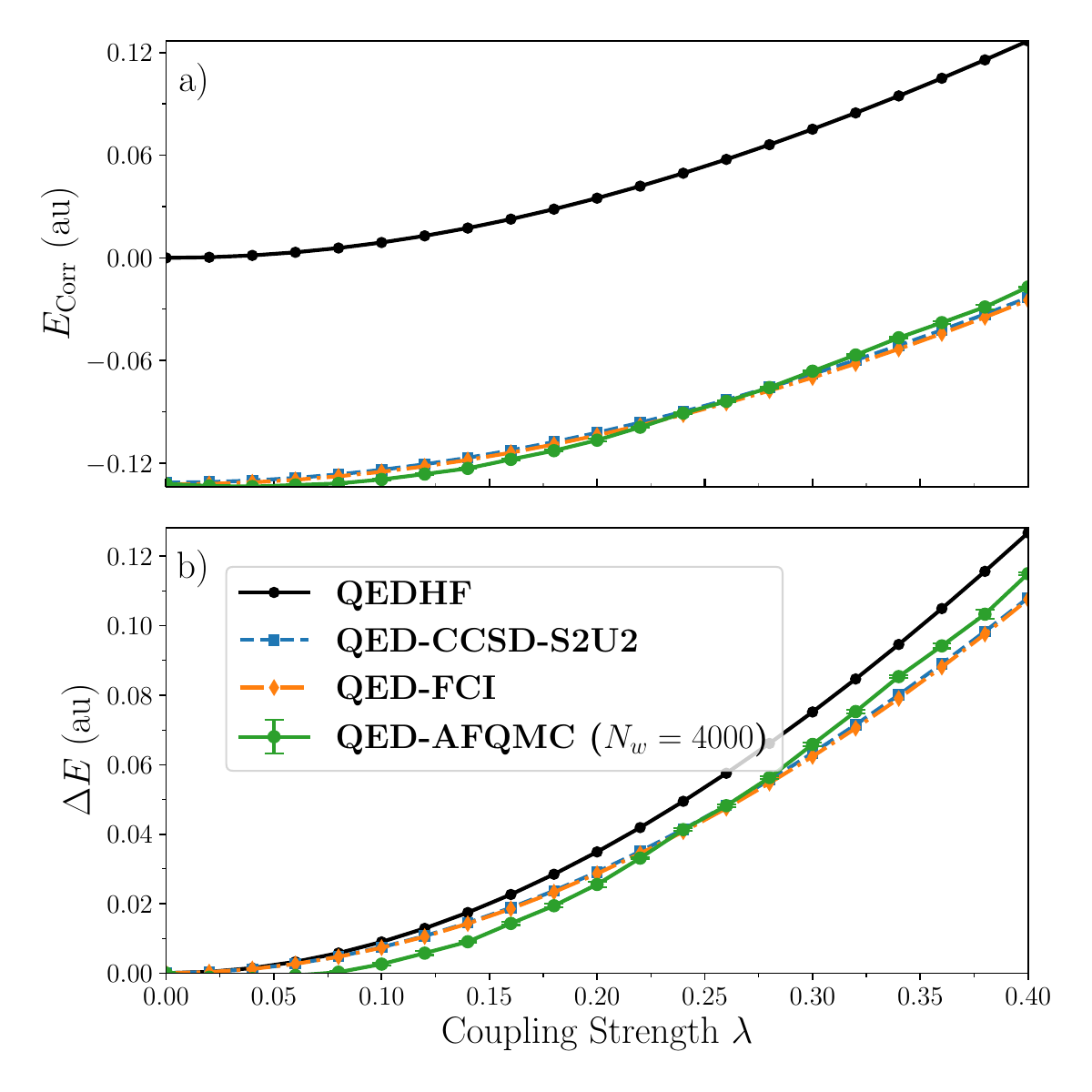}
    \caption{a) Correlation energies and b) DSE-mediated changes in correlation energies of the HF molecule as a function of electron-photon coupling strength $\lambda$ (with all terms of the PF Hamiltonian) are compared using QED-CCSD, QED-FCI, and QED-AFQMC.}
    \label{fig3}
\end{figure}

Next, we include all terms of the PF Hamiltonian and compare the total energies computed using several QED electronic structure methods, including QED-HF, QED-CCSD-S2U2 (which includes electronic, photonic, and electron-photon excitations up to second order), QED-FCI, and QED-AFQMC. The results are shown in Figure~\ref{fig3}. 
To enable a fair comparison between methods, we first analyze results obtained with a photonic Fock state truncation at $n=2$ for QED-CCSD-S2U2, QED-FCI, and QED-AFQMC (see Supplementary Figure~\ref{figs1}). In this case, QED-CCSD-S2U2 exhibits notable discrepancies in the strong coupling limit due to its underestimation of electronic correlation, as previously demonstrated in Figure~\ref{fig2}. This leads to a different topology in the energy landscape relative to QED-FCI and QED-AFQMC.
Figure~\ref{fig3} then shows results using a higher photonic Fock truncation at $n=5$ for QED-AFQMC and QED-FCI methods, where we observe a surprisingly good agreement between QED-CCSD-S2U2 and QED-FCI. However, this agreement arises from a cancellation of errors: QED-CCSD-S2U2 tends to underestimate both the electronic correlation (which would lower the correlation energy) and the electron–photon correlation (which would increase the energy), resulting in an artificial balance that masks the true discrepancies.
Crucially, across all tested truncation levels, QED-AFQMC maintains close agreement with QED-FCI, reflecting its ability to capture both electronic and photonic correlations accurately. Unlike QED-CCSD and QED-FCI, where increasing the number of photon excitations significantly raises computational cost, QED-AFQMC benefits from the decoupled propagation of the electronic and photonic degrees of freedom, enabling efficient treatment of large photonic Fock states without additional computational overhead (compared to the electronic parts). 
These results highlight QED-AFQMC as a powerful and scalable approach for accurately modeling strongly coupled light–matter systems.

\begin{figure}
    \centering
    \includegraphics[width=0.99\linewidth]{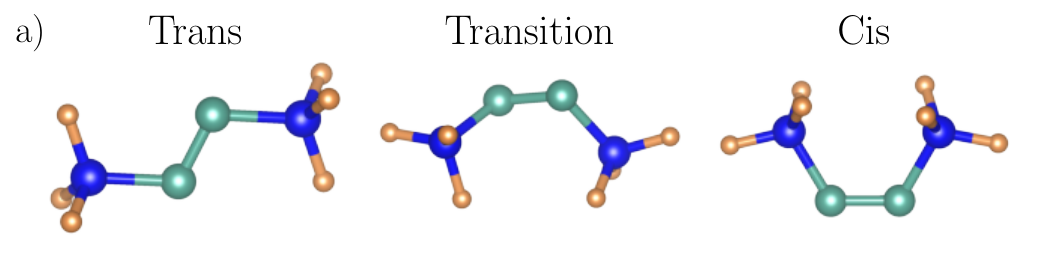}
    \includegraphics[width=0.99\linewidth]{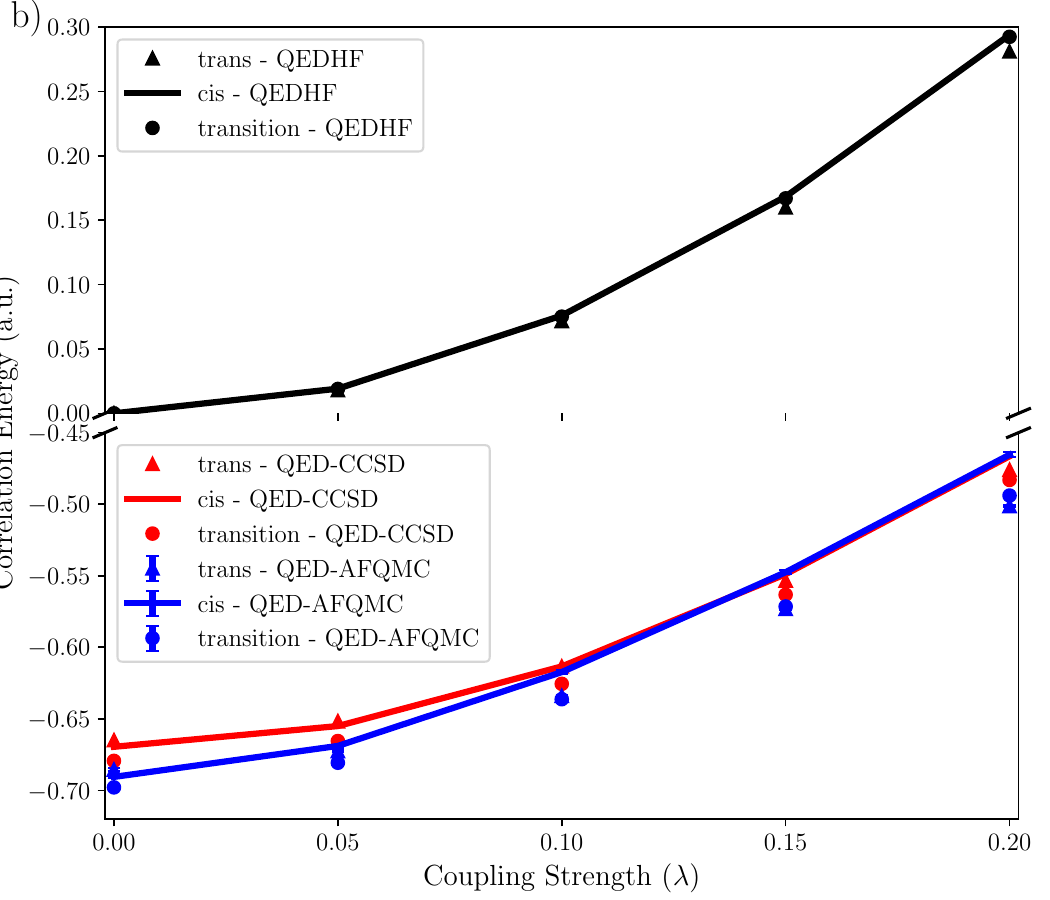}
    \includegraphics[width=0.99\linewidth]{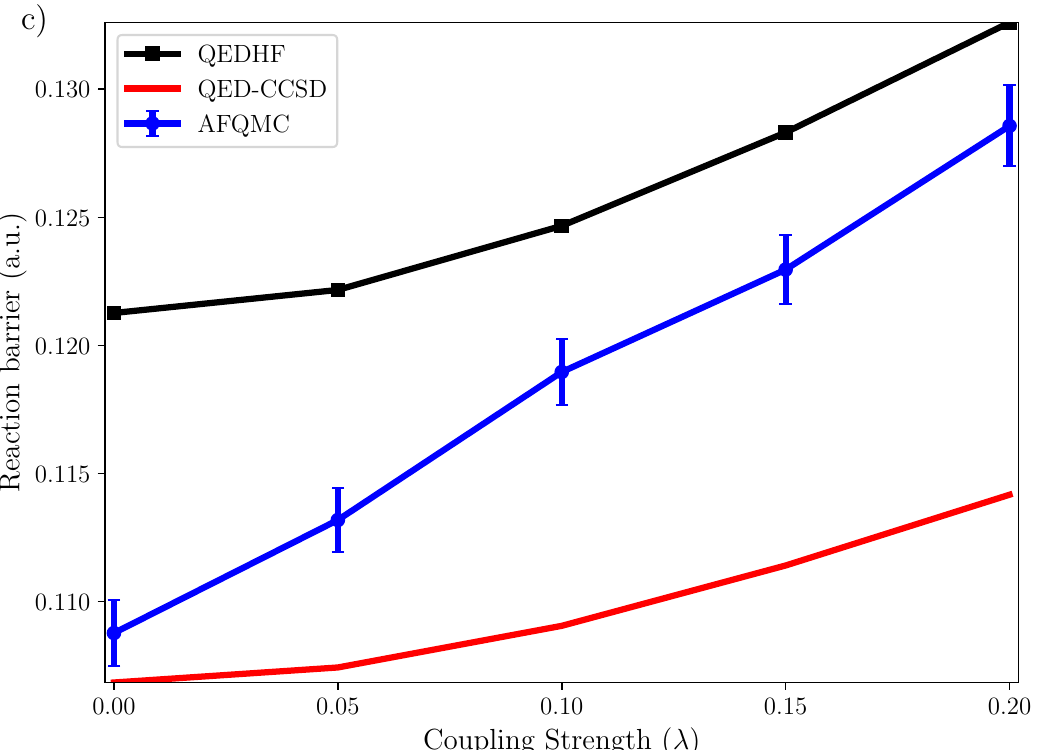}
    \caption{(a) \textit{Trans}, \textit{transition}, and \textit{cis} configurations of the C\textsubscript{2}N\textsubscript{2}H\textsubscript{6} isomer. (b) Correlation energies, and (c) reaction barrier (between the trans and transition configurations),  as functions of the coupling strength for the C\textsubscript{2}N\textsubscript{2}H\textsubscript{6} isomer.}
    \label{fig5}
\end{figure}

Moreover, we demonstrate the developed QED-AFQMC method for studying cavity effects in chemical transformations using the C\textsubscript{2}N\textsubscript{2}H\textsubscript{6} isomer~\cite{Zhang2019JCP} as an example. 
Figure~\ref{fig5} presents a detailed comparison of the isomerization reaction in the C\textsubscript{2}N\textsubscript{2}H\textsubscript{6} molecule under varying light-matter coupling strengths. Panel (a) shows the molecular configurations corresponding to the \textit{trans}, \textit{transition}, and \textit{cis} states. Panel (b) plots the correlation energies for each configuration, where triangles, lines, and solid circles denote the \textit{trans}, \textit{cis}, and \textit{transition} states, respectively. The black, red, and blue curves correspond to QED-HF, QED-CCSD-S2U2, and QED-AFQMC results. Panel (c) displays the reaction barrier, defined as the energy difference between the \textit{transition} and \textit{trans} configurations, for each method. The shaded region in the AFQMC curve indicates the standard deviation of the barrier height, calculated as $\sigma=\sqrt{\sigma_{\text{trans}}^2+\sigma_{\text{transition}}^2}$
where $\sigma_{\text{trans}}$ and $\sigma_{\text{transition}}$ are the standard deviations of the AFQMC correlation energies for the respective geometries. From these results, we observe that while capturing the qualitative trend, QED-HF significantly overestimates the correlation energy due to its neglect of electronic correlation and overestimation of electron-photon correlation. This leads to an overprediction of the cavity-induced stabilization and, consequently, an overestimated reaction barrier. In contrast, QED-CCSD includes electronic correlation but tends to underestimate the total correlation energy and the effect of light-matter coupling, leading to lower predicted barriers. QED-AFQMC, which treats both electronic and photonic correlations at higher truncation, yields intermediate values that likely offer a more accurate estimate of the cavity-modified reactivity. These findings underscore the limitations of mean-field and truncated QED-CCSD method at lower order and highlight the importance of fully correlated treatments—like QED-AFQMC—for reliably predicting chemical behavior in strongly coupled light-matter systems.

\section{Discussion and Summary}
\label{sec: summary}
In this work, we developed a QED-AFQMC method in second quantization for solving the Pauli-Fierz Hamiltonian in the context of molecular systems coupled to quantized radiation fields. By casting the coupled electron-photon Hamiltonian into a Monte Carlo-friendly form via a combination of modified Cholesky decomposition and tailored Hubbard-Stratonovich transformations, we enabled a unified treatment of fermionic and bosonic degrees of freedom in strongly coupled light-matter systems.
We validated the QED-AFQMC approach by comparing against exact full configuration interaction (QED-FCI) and coupled-cluster (QED-CCSD) benchmarks for small systems like HF, demonstrating that QED-AFQMC accurately captures correlation effects across a wide range of coupling strengths. While QED-CCSD systematically underestimates the electron-photon correlation energy, especially in the strong coupling regime, QED-AFQMC closely matches FCI results and retains numerical stability and scalability, making it a promising approach for studying polaritonic ground states beyond the reach of conventional methods.

We further applied QED-AFQMC to the C\textsubscript{2}N\textsubscript{2}H\textsubscript{6} isomerization reaction to explore cavity-modified reactivity. Our analysis shows that QED-HF tends to overestimate the effect of the cavity due to the lack of electronic correlation and overestimation of electron-photon contributions, while QED-CCSD underestimates the reactivity changes due to the truncation of correlation effects. QED-AFQMC provides a balanced description, yielding reaction barrier predictions between QED-HF and QED-CCSD, offering a more accurate estimate of cavity-modified chemical transformations. These findings highlight the necessity of non-perturbative and systematically improvable approaches for accurately modeling quantum optical effects in molecular chemistry.

Our implementation combines C++ and Python, enhancing both portability and computational performance. The C++ backend ensures efficient numerical routines and scalability, while the Python interface provides flexibility for rapid prototyping and integration with existing electronic structure packages.
Our modular design also makes it straightforward to study other custom model Hamiltonians, such as Hubbard–Holstein models~\cite{Costa:2020aa}, by simply modifying the operator definitions within the AFQMC framework.
Looking forward, an important next step is the integration of our QED-AFQMC formalism with variational photonic wavefunction ansätze, particularly the variational displacement~\cite{li_vQED_arXiv2023} and squeezing ans\"{a}tze~\cite{Mazin:2024aa} we have recently developed. We anticipate that this hybrid framework will significantly enhance the numerical efficiency of QED-AFQMC by reducing the number of required Fock states in the photonic Hilbert space and improving convergence along the imaginary time propagation. Such an approach could provide a path toward chemically accurate, scalable modeling of complex cavity-modified phenomena in realistic molecular and material systems. Overall, QED-AFQMC offers a robust, accurate, and extensible platform for studying correlated light-matter systems and lays the groundwork for future developments in quantum chemistry under strong light-matter interaction regimes.

\section*{Code Availability Statement} The method presented in this work is implemented in the open-source package OpenMS~\cite{openms}, which is available at \url{https://github.com/lanl/OpenMS}.

\begin{acknowledgments}
We acknowledge support from the US DOE, Office of Science, Basic Energy Sciences, Chemical Sciences, Geosciences, and Biosciences Division under Triad National Security, LLC (``Triad'') contract Grant 89233218CNA000001 (FWP: LANLECF7). This research used computational resources provided by the Institutional Computing (IC) Program and the Darwin testbed at Los Alamos National Laboratory (LANL), funded by the Computational Systems and Software Environments subprogram of LANL's Advanced Simulation and Computing program. LANL is operated by Triad National Security, LLC, for the National Nuclear Security Administration of the US Department of Energy (Contract No. 89233218CNA000001).
\end{acknowledgments}

\appendix

\section{Additional figure}
\begin{figure}[h]
    \centering
    \includegraphics[width=0.99\linewidth]{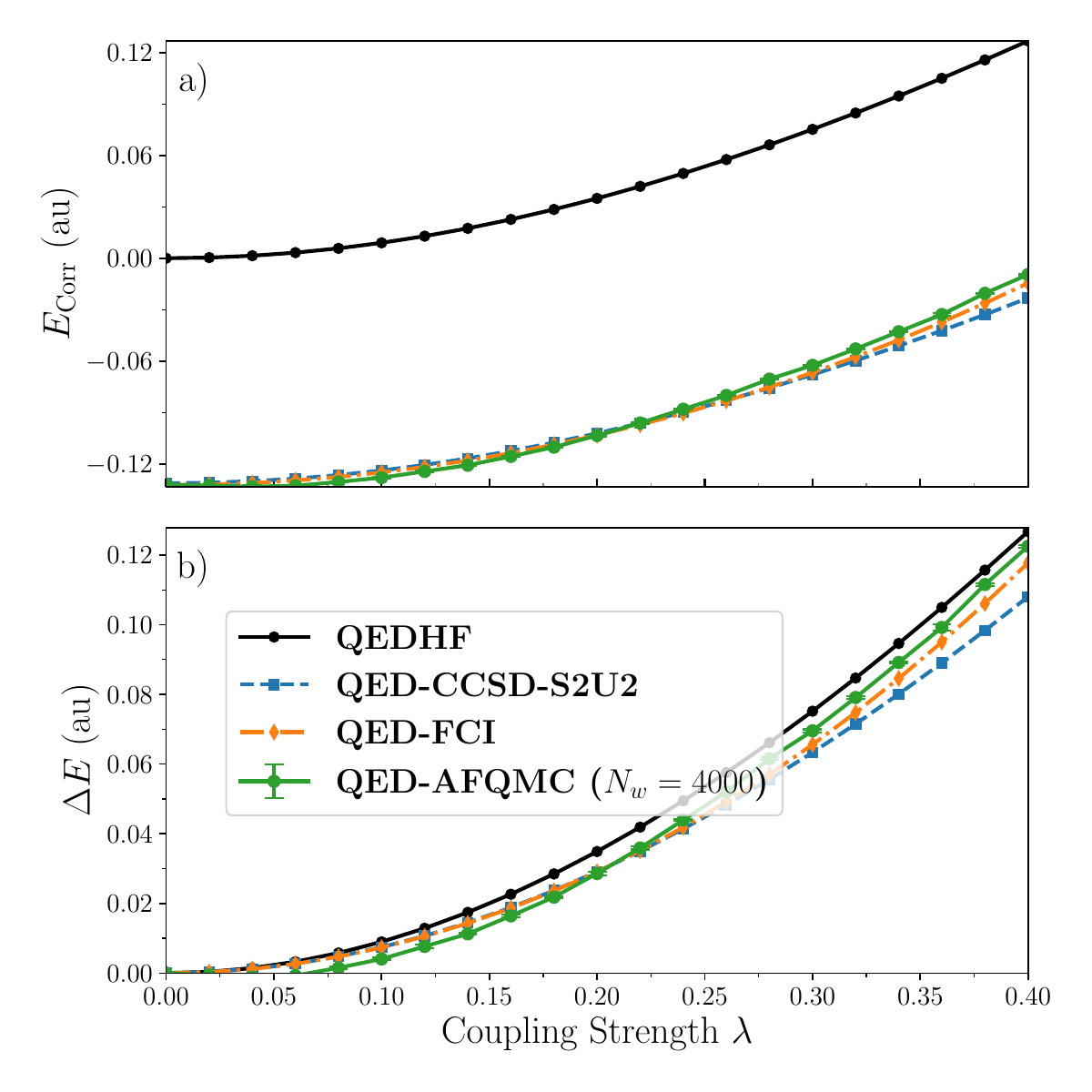}
    \caption{Same plots as in Figure~\ref{fig3}, but with the photonic Fock space truncated at $n=2$.}
    \label{figs1}
\end{figure}

\bibliography{qmc}

\end{document}